# Quasi-static magnetic compression of field-reversed configuration plasma: Amended scalings and limits from two-dimensional MHD equilibrium

Abba Alhaji Bala [1,2,5], Ping Zhu [1,3], Haolong Li [4], Yonghua Ding [1], Jiaxing Liu [1], Sui Wan [1], Ying He [1], Da Li [1], Nengchao Wang [1], Bo Rao [1], and Zhijiang Wang [1]

[1] International Joint Research Laboratory of Magnetic Confinement Fusion and Plasma Physics, State Key Laboratory of Advanced Electromagnetic Engineering and Technology, School of Electrical and Electronic Engineering, Huazhong University of Science and Technology, Wuhan 430074, People's Republic of China
[2] School of Physics, Huazhong University of Science and Technology, Wuhan, Hubei 430074, China
[3] Department of Engineering Physics, University of Wisconsin-Madison, Madison, Wisconsin 53706,United States of America
[4] College of Physics and Optoelectronic Engineering, Shenzhen University, Shenzhen, 518060, China
[5] Department of Physics, Federal University Dutse, Jigawa, Nigeria

E-mail: zhup@hust.edu.cn, yhding@hust.edu.cn

## Abstract

In this work, several key scaling laws of the quasi-static magnetic compression of field reversed configuration (FRC) plasma [Spencer, Tuszewski, and Linford, 1983] are amended from a series of 2D FRC MHD equilibriums numerically obtained using the Grad-Shafranov equation solver NIMEQ. Based on the new scaling for the elongation and the magnetic fields at the separatrix and the wall, the empirically stable limits for the compression ratio, the fusion gain, and the neutron yield are evaluated, which may serve as a more accurate estimate for the upper ceiling of performance from the magnetic compression of FRC plasma as a potential fusion energy as well as neutron source devices.

## 1. Introduction

Field-reversed configuration (FRC) is an elongated compact torus plasma sustained solely by a poloidal magnetic field [1]. Because the FRC has no or little toroidal magnetic field, it has a very high plasma beta ($< \beta >$ averaged over the separatrix volume lies in the range 0.5 - 1). The magnetic field of FRC is composed of open magnetic field lines, separatrix, and closed magnetic field lines, while the plasma almost entirely exists inside the separatrix with an attractive feature of having no material objects linking the torus. The two FRC regions are referred to as "open" and "closed" respectively [2].

FRC has been one of the preferred candidate configurations for fusion devices such as compact nuclear fusion reactors and neutron sources, which has received increasing attentions from various countries. Both experimental and theoretical studies have been performed to explore the potentials and challenges associated with such an attractive path towards fusion [3–6]. The C-2 experiment in TAE Technology has been able to obtain and maintain an FRC with $n$ reaching $3 \times 10^{19} \, m^{-3}$, $T_e$ and $T_i$ around $1 keV$ respectively, the energy confinement time $\tau_E$ about $1ms$, and the $\beta$ around 90 percent [7]. The LANL and the ALPHA project in the United States successfully increased the temperature and density of an FRC plasma by 1 order of magnitude using staged magnetic compression [6]. Japan, Canada, Russia and other countries also have related experimental devices as well [8]. In China, the Huazhong Field Reversed Configuration (HFRC) has been designed to explore a novel concept of "two-staged" magnetic compression of FRC as a path to achieving a compact and economic neutron source and potential fusion reactor [9].

Magnetic compression of FRC is one of the promising paths to fusion that has been pursued over years. For instance, the Princeton ATC device [10] have increases the plasma density by 5 times with adiabatic compression. The FRX-C/LSM device of Los Alamos National Laboratory also adopted the adiabatic compression method for FRC, and the plasma temperature and density are increased by 10 times and 5 times respectively [11].

A one-dimensional model for the adiabatic compression of FRC [12,13] was established. In this theory, the quasi-static approximation is used such that compression process is considered as a series of MHD equilibria with sequentially varying compression ratio. Although the magnetic compression of FRC plasma is a highly nonlinear 3D dynamic process, the quasi-static approximation allows us to establish the scaling laws and to evaluate the upper stable limits achievable for the compression ratio, fusion gain and neutron yields from such an approach.

However, 1D approximations are often made in order to obtain the analytical scaling laws for compression, where the 2D MHD equilibrium conditions or constraints on the FRC parameters are often not well satisfied. The FRC equilibrium is essentially two-dimensional, and previous theory models for FRC magnetic compression often fail to take into account of the two-dimensional spatial and geometric features of FRC equilibrium. More importantly, previous FRC magnetic compression theories have not considered the constraints imposed on the accessible parameter space due to the macroscopic instabilities of FRC plasmas. Therefore, subject to the constraints from the strict FRC two-dimensional equilibrium and the stability criterion, whether the FRC plasma parameters can achieve the fusion ignition conditions through the approach of magnetic compression, remains one of the primary problems to address in the design of FRC neutron source and fusion reactor today.

In order to explore this major issue, in this work, we use a series of two-dimensional FRC MHD equilibria from numerically solutions of the Grad-Shafranov equation to obtain the amended scaling laws for the key parameters of the FRC plasma during quasi-static magnetic compression including two-dimensional spatial geometric effects, and together with empirical criterion for FRC kinetic MHD stability, to evaluate the fusion ignition conditions and the upper limit of neutron yield that can be achieved through the stable FRC magnetic compression process in the resistive MHD model.

The rest of this work is organized as follows. In section 2, the numerical FRC equilibrium solution is determined and checked for convergence. In section 3, the 2D MHD equilibriums during a quasi-static FRC compression process are solved and the numerical solutions are used to obtain the amended scalings for compression. In section 4, the empirically stable limits for the compression ratio, the fusion gain, and the neutron yield are evaluated. Finally, summary and discussion are given in section 5.

**2. The FRC equilibrium**

In this work, we solve for the 2D MHD equilibrium of FRC that are consistent with the analytical scaling law for the maximum pressure in radial profile during the magnetic compression based on a 1D approximation. The 2D MHD equilibrium of FRC can be obtained from solving the Grad–Shafranov equation (GSE). Besides the limited analytical GSE solutions of FRC equilibria [14–17], numerical methods have been applied to solving the GS equation using 2-D spectral element [18], method of fundamental solutions (MFS), finite difference method, boundary element method (BEM), conformal mapping and Green's function. Correspondingly, numerical toroidal equilibrium codes have been developed, such as EFIT, CHEASE, ESC, and NIMEQ [15]. In this study, we use the Grad-Shafranov equilibrium solver NIMEQ which is based on the spectral element expansions in two dimensions. The NIMEQ solutions for the 2D FRC equilibriums with different compression ratios are then used to obtain the compression scaling laws, which are compared with the Spencer theory based on 1D approximation.

From the force balance equation, the magnetic divergence constraint, and Ampere's law, we obtain the Grad-Shafranov equation for the MHD equilibria of an axisymmetric toroidal system. The FRC is characterized by a zero toroidal field, thus the GS equation for the FRC equilibrium takes the form

$$\Delta^*\psi = \mu_0 R^2 \frac{dP}{d\psi}, \qquad (1)$$

where $\psi$ is the magnetic flux function, $P(\psi)$ is the plasma pressure and $\Delta^* = R\frac{\partial}{\partial R}R^{-1}\frac{\partial}{\partial R} + \frac{\partial^2}{\partial Z^2}$ is the Grad-Shafranov operator. For a quadratic pressure profile

$$P = P_0 + P_2(\psi^2/\psi_t^2) \qquad (2)$$

where $\psi_t$ is the flux inside the separatrix, $P_0$ and $P_2$ are constants, Eq. (1) is reduced to the following linear equation for $\psi$

$$\Delta^*\psi = 2\mu_0 R^2 P_2\left(\frac{\psi}{\psi_t^2}\right). \qquad (3)$$

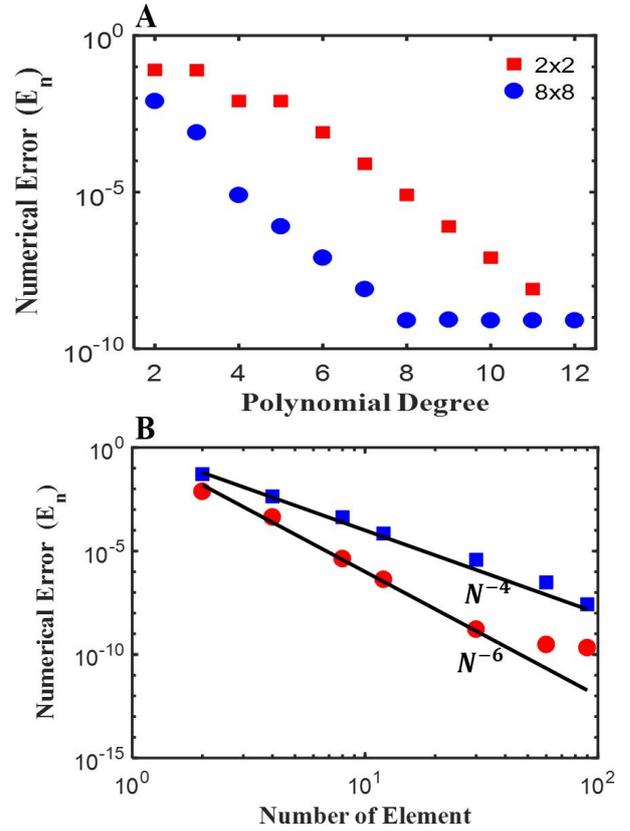

**Figure 1.** (A) The numerical error $E_n$ of $\psi$ as a function of the element polynomial degree for $2 \times 2$ element mesh (■) and 8x8 element mesh (●) in FRC equilibrium. (B) The numerical error $E_n$ of $\psi$ as a function of the number of elements for the 2nd order elements (■) and 4th order elements (●). The black lines stand for the scaling fitted from the decaying numerical errors, where N denotes the number of elements.

We look for solutions in the cylindrical coordinates R and Z that are symmetric with respect to the middle plane $Z = 0$. With boundary conditions $\psi(R, Z) = 0$ at $R = a$ and $Z = \pm b$, where $h = 2b$, and $a$ and $h$ are the radius and the length of the cylinder containing the FRC plasma respectively, we obtain

$$\psi(R, Z) = \psi_0 \frac{F_0\left(\eta, \frac{\sqrt{d}R^2}{2}\right)}{F_0\left(\eta, \frac{\sqrt{d}R_0^2}{2}\right)} \cos(\lambda Z), \quad (4)$$

where $F_0$ is the zeroth-order Coulomb wave function, $d = 2\mu_0 P_2 / \psi_0^2$, $\eta = \lambda^2/(4\sqrt{d})$ and $\lambda = \pi/h$.

For a more general pressure profile $P(\psi)$, we numerically solve the GS equation Eq. (1) using the NIMEQ code, which is a Grad-Shafranov equilibrium solver developed within the framework of NIMROD for the more realistic geometry [18]. The finite element method is used to solve the GS equation and the Picard scheme is used to advance the iteration. To demonstrate the numerical accuracy and convergence of the NIMEQ code, the numerical and analytical solutions of Eq. (1) are compared for the special case of pressure profile in Eq. (2) in terms of the numerical error defined by

$$E_n = \sqrt{\sum (\psi_N - \psi_A)^2 / \sum \psi_A^2} \quad (5)$$

where $\psi_N$ is the numerical solution from NIMEQ and $\psi_A$ is the analytic solution from Eq. (4), and the summation is performed over all of the finite-element nodes.

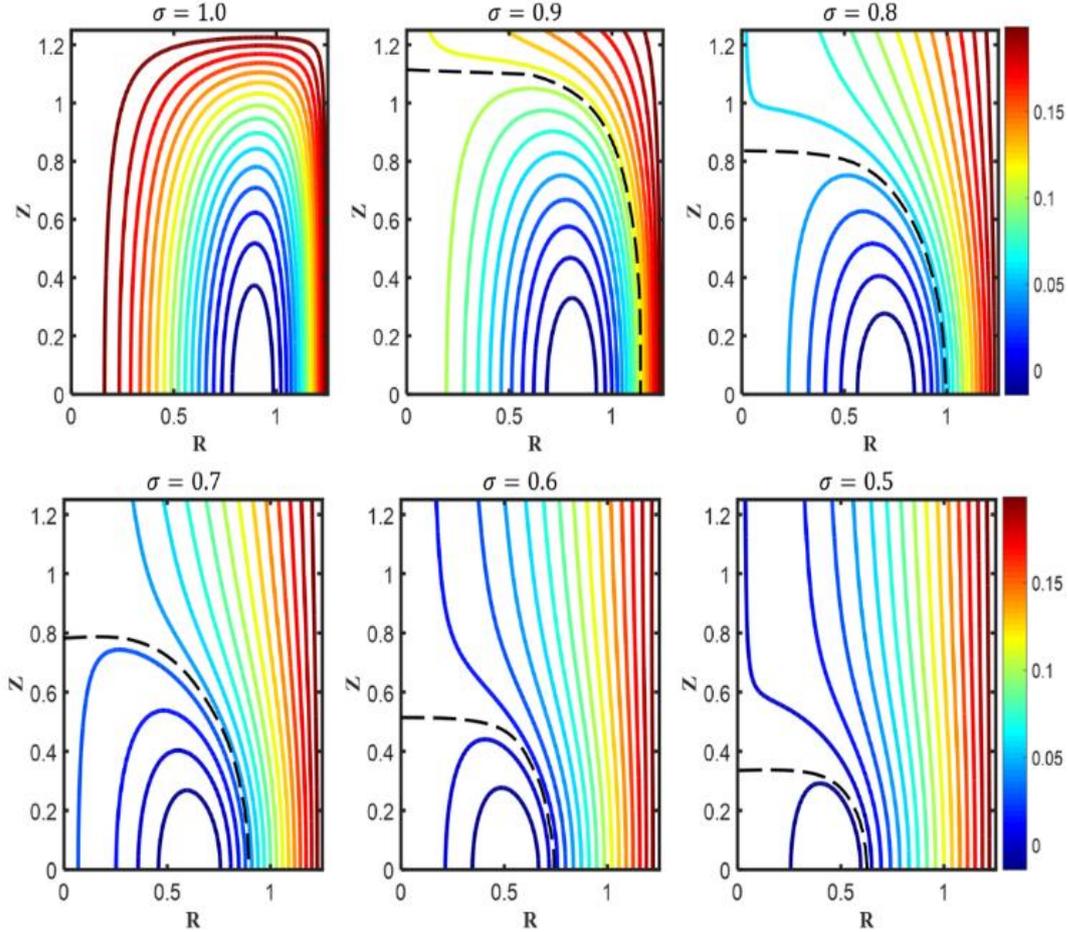

**Figure 2.** The contour of magnetic fluxes of 2D MHD equilibrium with initial elongation $e_0 = 2$ and $P_{mi} = 0.0018$.

Two methods are applied to examine the convergence of the NIMEQ solution, namely, the h-refinement and p-refinement. In the p-refinement method, the polynomial degree is increased as the number of elements is kept constant. In contrast, in the h-refinement method, the polynomial degree is fixed while the number of elements is varied. Both methods of comparisons show numerical convergence of the NIMEQ solutions (Fig. 1).

## 3. FRC equilibrium during quasi-static magnetic compression

The pressure profile inside the separatrix of an FRC plasma during compression can be modelled [12] as

$$P(\psi) = P_m \beta(\phi) \quad (6)$$

where $\beta(\phi) = (\phi + \sigma\phi^2)/(1+\sigma)$ and $\phi = \psi/\psi_t$. $P_m$ is the maximum value of pressure profile, $\sigma$ is the compression ratio, $\psi$ is the poloidal flux, and $\psi_t$ is the magnetic flux inside the separatrix with $\psi_t = \psi_{separatrix} - \psi_{axis}$. The compression ratio is defined as $\sigma = R_s/R_W$, for which $R_s$ is the radius of the separatrix and $R_W$ is the wall major radius in the middle plane at $Z = 0$. Applying the adiabatic compression scaling law based on 1D approximation [12], we have

$$P_m = P_{mi}\sigma^{-2(3-\epsilon)}, \quad (7)$$

where $P_{mi}$ is the initial maximum pressure value before compression, and $\epsilon = -0.25$ for a typical high flux sharp-boundary case.

Let $\Psi = \frac{\psi}{\psi_t} + \frac{1}{2\sigma}$, thus the G-S equation for the FRC plasma inside the separatrix becomes

$$\Delta^*\Psi = -\mu_0 R^2 \frac{2P_m\sigma}{(1+\sigma)\psi_t^2}\Psi = -dR^2\Psi, \quad (8)$$

where $d = \frac{2\mu_0 P_m \sigma}{(1+\sigma)\psi_t^2}$.

For any specific radial compression ratio, the 2D FRC equilibrium equation (8) is numerically solved using NIMEQ for the set of the external coils that are so chosen to ensure that the radial compression ratio $R_s/R_w$ at $Z = 0$ from the NIMEQ solution matches the $\sigma$ in the pressure profile (Figs. 2 and 3). For both small and large initial elongations, the last closed flux surface as well as the separatrix of FRC shrinks quickly along the axial direction (i.e. Z-direction) as the FRC is compressed radially to each value of $\sigma$. The FRC length $l$, the elongation $e$, and the magnetic field at the wall $B_w$ are compared between the measured values from the 2D MHD equilibrium solutions and the following Spencer scaling law (Eqs. 9 - 11) for various radial compression ratios (Fig. 4),

$$l_s \propto \sigma^{2(3-\epsilon-\gamma)/\gamma}\left(1-\frac{\sigma^2}{2}\right)^{-\frac{1+\epsilon-\gamma\epsilon}{\gamma}}, \quad (9)$$

$$e = \frac{l_s}{R_s} \propto \frac{\sigma^{2(3-\epsilon-\gamma)/\gamma}\left(1-\frac{\sigma^2}{2}\right)^{-(1+\epsilon-\gamma\epsilon)/\gamma}}{\sigma R_{s0}}, \quad (10)$$

$$B_w \propto \sigma^{-3+\epsilon}, \quad (11)$$

where the length of FRC $l_s$ can be measured from the intersection between the separatrix and the Z-axis at $R = 0$, $R_{s0} = R_w$, and the magnetic field magnitude $B_w$ at wall is measured from the equilibrium solution at $(R_w, 0)$.

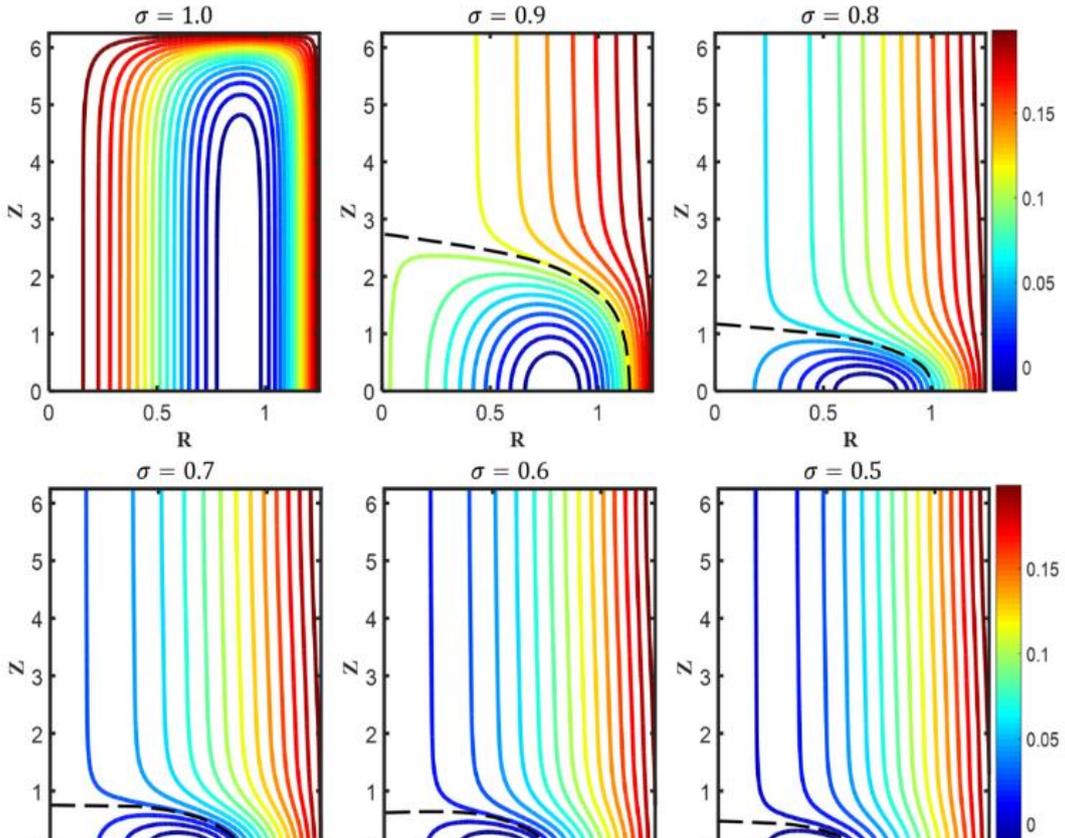

**Figure 3.** The contour of magnetic fluxes of 2D MHD equilibrium with initial elongation $e_0 = 10$ and $P_{mi} = 0.0018$.

Whereas the comparisons show overall quantitative agreement, the degree of quantitative agreement depends on the FRC equilibrium elongation, which is defined as the ratio of the FRC length $l_s$ over radius $R_s$ at separatrix for the initial elongation $e_0 = 3$, the agreement on the $l_s$ and $e$ scalings are the best for the smaller initial elongation, the FRC shape measured from 2D equilibrium shrinks slower than the 1D Spencer scaling law. As the initial elongation increases, the shrinking of FRC during compression becomes faster from 2D equilibrium calculations than 1D Spencer scaling law, as indicated in Figs. 4A and 4B. This finding may appear surprising, since the Spencer scaling law is expected to apply best when FRC is highly elongated and hence the 1D approximation is more valid. However, the 2D equilibrium calculations show that, even for the FRC with large initial elongation (e.g. $e_0 \geq 10$), once the magnetic compression process begins, the FRC elongation itself quickly drops out of the regime where the 1D approximation is valid.

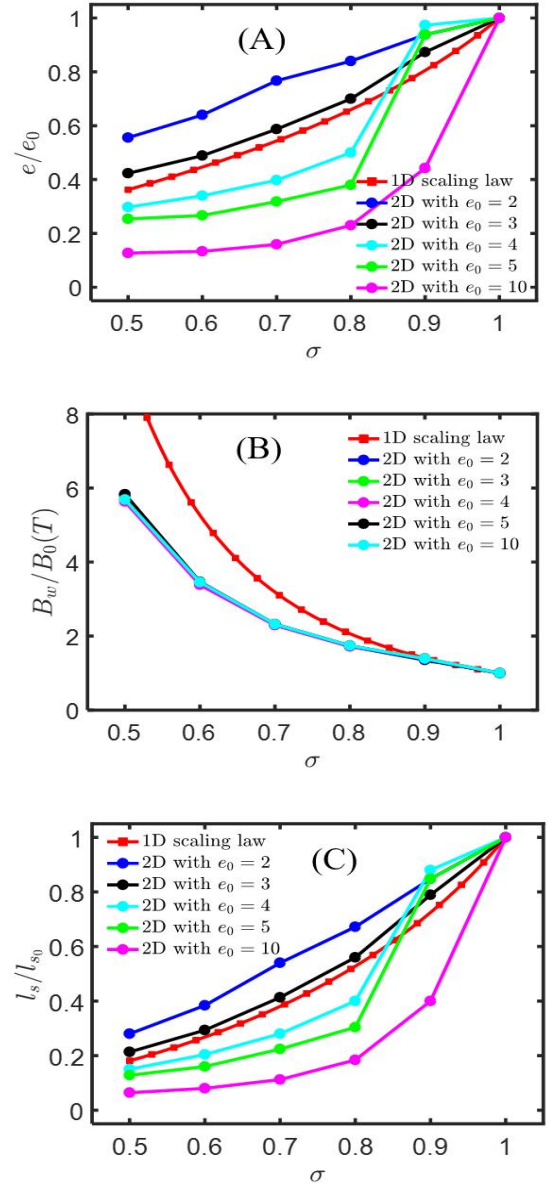

**Figure 4.** (A) Elongation $e$, (B) wall magnetic field $B_w$, and (C) FRC length $l_s$ as functions of the radial compression ratio $\sigma$ from 1D scaling law and 2D MHD equilibriums with various initial elongations.

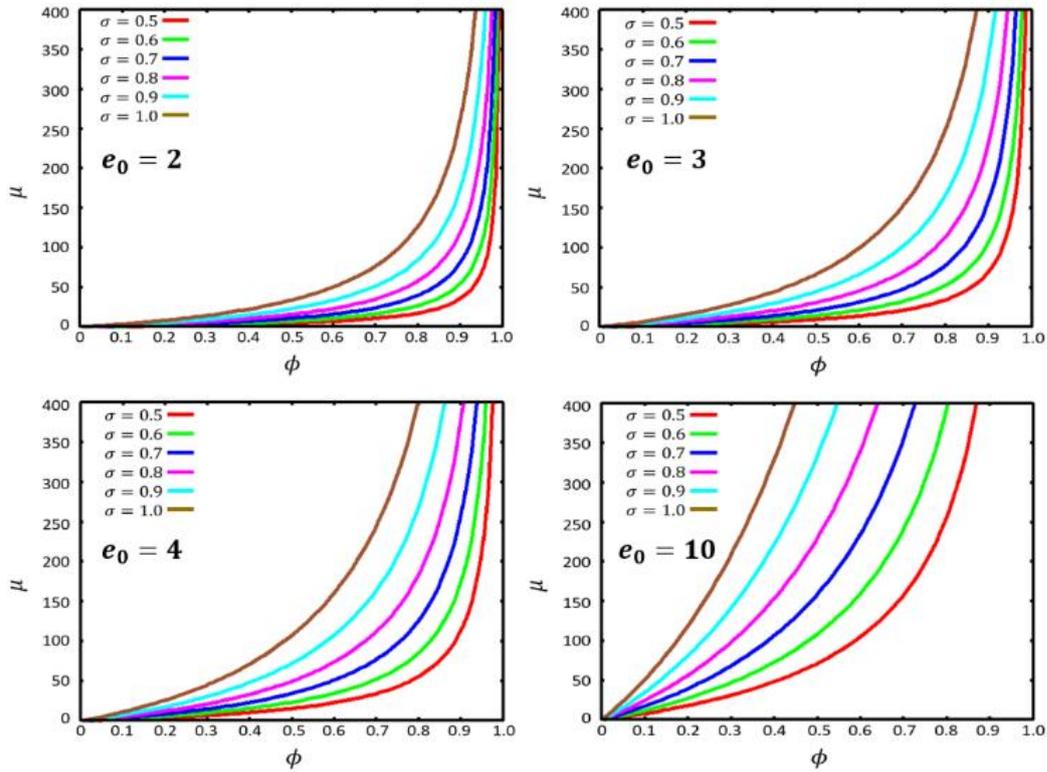

**Figure 5.** The $\mu(\phi)$ profiles calculated from 1D scaling laws with various initial elongations for $\mu_0 P_{mi} = 0.0018$.

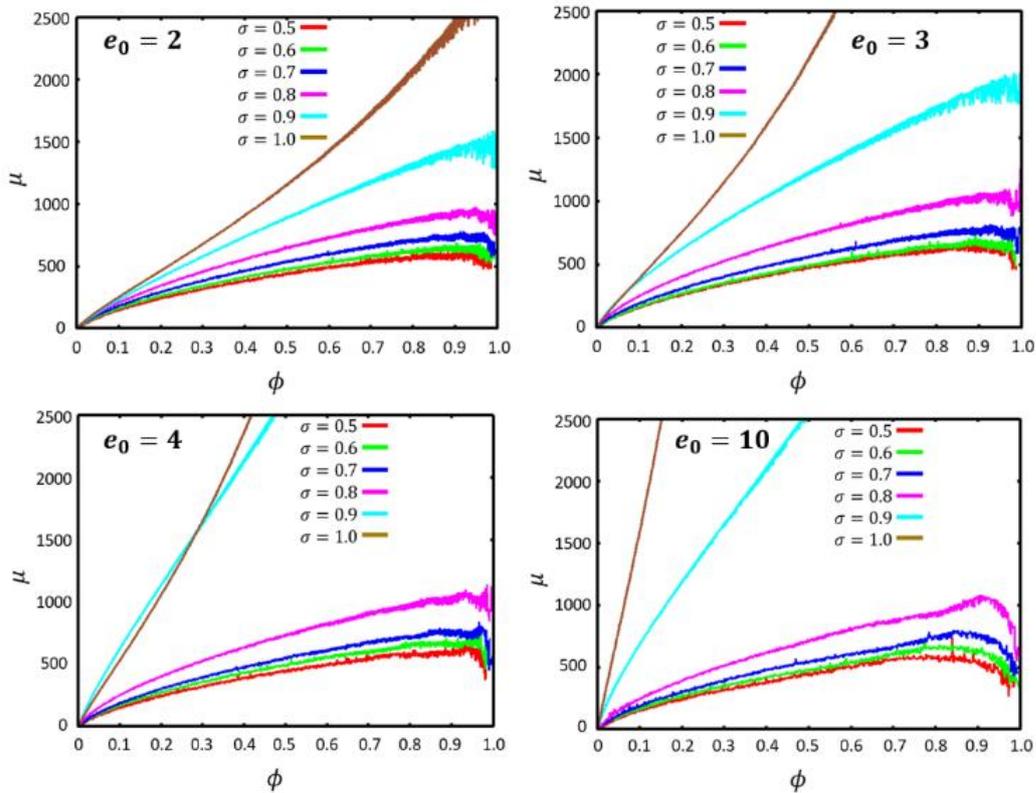

**Figure 6.** The $\mu(\phi)$ profiles evaluated from 2D MHD equilibriums with various initial elongations for $\mu_0 P_{mi} = 0.0018$.

For the entropy per unit flux

$$\mu(\psi) = P(\psi)\left(2\pi \oint \frac{dl}{B}\right)^\gamma, \quad (12)$$

its one-dimensional approximation for the elongated FRC is based on [12]

$$\oint \frac{dl}{B} \approx \frac{2l_s}{B} = \frac{2l_s}{\{2[P_m - P(\psi)]\}^{1/2}}, \quad (13)$$

where $l_s$ is the length of FRC separatrix. Comparison of the $\mu(\psi)$ profiles calculated from the 1D approximation using Eq. (13) (Fig. 5) and the 2D MHD equilibrium using Eq. (12) with various initial elongations and radial compression ratios (Fig. 6) suggest that the adiabatic condition underlying the Spencer scaling law is less satisfied in the quasi-static magnetic compression process modelled by the series of 2D MHD equilibriums. This may also contributes to the differences between compression scalings measured from the 2D equilibriums and calculated from the 1D Spencer scaling laws.

**4. Stable limits of FRC compression**

4.1. FRC stability criterion

Based on experimental data, an empirical stability criterion of FRC can be written as [19–21]

$$S/\kappa < 3.5, \quad (14)$$

where $\kappa = Z_s/R_s$ is the elongation of the separatrix and $S = R_s/\delta_i$ is the ratio of the radius of the separatrix to the ion skin depth: $\delta_i = c/\omega_{pi}$ and $\omega_{pi} = \sqrt{n_i e^2/\epsilon_0 m_i}$. From these we have

$$S/\kappa = \frac{R_s/\delta_i}{Z_s/R_s} = \frac{R_s^2}{Z_s c}\sqrt{\frac{n_i e^2}{\epsilon_0 m_i}}. \quad (15)$$

Using the analytical scaling law for the adiabatic compression of an elongated FRC [12],

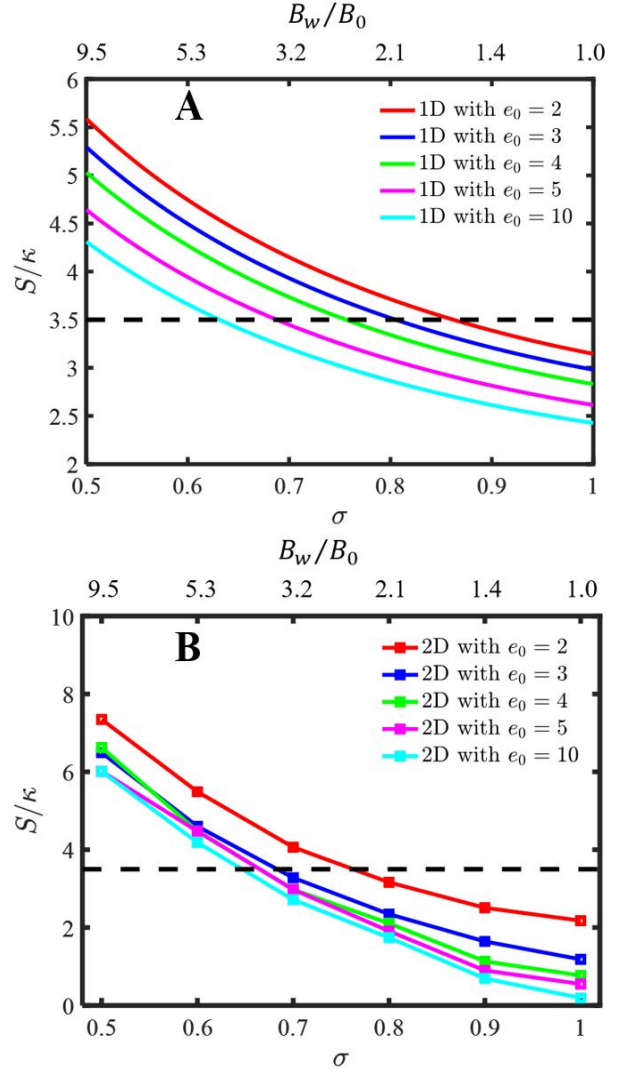

**Figure 7.** The $S/\kappa$ stability criterion calculated from (A) 1D scaling laws and (B) 2D MHD equilibrium for different radial compression ratio $\sigma$ and magnetic field strength at the wall $B_w$, with various initial elongations, $n_{i0} = 3.0 \times 10^{19} m^{-3}$, $\gamma = 5/3$, and $\epsilon = -1/4$.

$$Z_s = \frac{1}{2}l_{s0}\sigma^{2(3-\epsilon-\gamma)/\gamma}\left(1 - \frac{\sigma^2}{2}\right)^{-(1+\epsilon-\gamma\epsilon)/\gamma}, \quad (16)$$

$$n_i = n_{i0}\sigma^{-2(3-\epsilon)/\gamma}\left(1 - \frac{\sigma^2}{2}\right)^{-(1+\epsilon)(\gamma-1)/\gamma}, \quad (17)$$

where $l_{s0}$ and $n_{i0}$ are the scale length of the separatrix and number density for the initial equilibrium, respectively, we obtain an estimate of the ratio $S/\kappa$ for a given radial compression ratio $\sigma$ as

$$S/\kappa = \frac{2\sigma^2 R_{s0}^2}{c l_{s0} \sigma^{2(3-\epsilon-\gamma)/\gamma}\left(1-\frac{\sigma^2}{2}\right)^{\frac{1+\epsilon-\gamma\epsilon}{\gamma}}}$$

$$\sqrt{\frac{e^2 n_{i0} \sigma^{-2(3-\epsilon)/\gamma}\left(1-\frac{\sigma^2}{2}\right)^{-(1+\epsilon)(\gamma-1)/\gamma}}{\epsilon_0 m_i}}. \quad (18)$$

We also evaluated the ratio $S/\kappa$ based on its definition in Eq. (15) directly using the 2D MHD equilibrium for each given $\sigma$ and the magnetic field strength at the wall $B_w$, where the FRC length is measured from the separatrix and the $n_i$ is calculated from $n_i = \frac{N_i}{V} = \frac{N_i}{\pi R_s^2 l_s} = \frac{N_i}{\pi \sigma^2 R_{s0}^2 l_s} = \frac{n_{i0} l_{s0}}{\sigma^2 l_s}$. Figure 7(A) and figure 7(B) compare the $S/\kappa$ values determined from 1D analytical scaling law and those directly measured from the 2D MHD equilibrium. The black dashed line signifies the boundary where $S/\kappa$ equals to 3.5. Below this line is the stable regime for FRC magnetic compression. The measured estimate for $S/\kappa$ from the exact 2D MHD equilibrium solutions amends the prediction from the 1D analytical scaling law about the stable FRC compression limit.

4.2. Stable ignition regime

The stability criterion $S/\kappa > 3.5$ sets another boundary for the ignition parameter regime through the magnetic compression of an FRC plasma. One such example is demonstrated in Fig. 8, where the Lawson criterion for DT and DD reaction [22,23]

$$nT\tau_E > 3 \times 10^{21} m^{-3} keV s, \quad (19)$$

$$nT\tau_E \geq 10^{23} m^{-3} keV s, \quad (20)$$

are evaluated in the parameter space $(T_{mi}, \sigma)$ and $(T_{mi}, B_w)$, where $n$ is the plasma density, $T$ is the plasma temperature, $\tau_E$ is the energy confinement time, $T_{mi}$ is the initial maximum ion temperature, $B_w$ is the magnetic field strength at the wall and $\sigma$ is the radial compression ratio.

Ideally, for the adiabatic compression process, the energy confinement time $\tau_E$ would be infinity. More realistically, we may take the resistive diffusion time of magnetic field as an estimate for the upper limit of $\tau_E$. Using the Spitzer model for resistivity [24], we thus have $\tau_E = \frac{\mu_0}{\eta_\perp}\left(\frac{R_s}{2}\right)^2$, where $\eta_\perp = 1.96 \times 2.8 \times 10^{-8} T_e^{-3/2}$ ohm m, $T_e = T_i = P_{mf}/n_i$ in keV and $R_s = \sigma R_w$. The upper limit for $\sigma$ and $B_w$ is determined from the stability condition $S/\kappa > 3.5$, along with 2D MHD equilibrium calculations as outlined in previous subsections.

Additionally, the amended scaling for FRC elongation during magnetic compression is applied to the estimate of the upper limits for the radial compression ratio along with the empirical stability criterion for FRC. The achievable stability regimes for fusion ignition and neutron yield rates from the approach of FRC compression are evaluated. The upper limits on the stable ignition regime from both D-D and D-T fusion reaction within the stable ignition window that is stable against global kinetically MHD modes is $10^{25} keV. s. m^{-3}$ at $T_{mi} = 1 keV$ and $\sigma = 0.5$ for 2D MHD calculation and $10^{26} keV. s. m^{-3}$ at $T_{mi} = 6 keV$ and $\sigma = 0.6$ for 1D calculation.

4.3. Limits of D-T and D-D neutron yield rates

We further estimate the upper limits on the neutron yield rates from D-T and D-D fusion reactions through FRC compression imposed by the empirical stability criterion. The neutron yield rate used herein is estimated using [25]

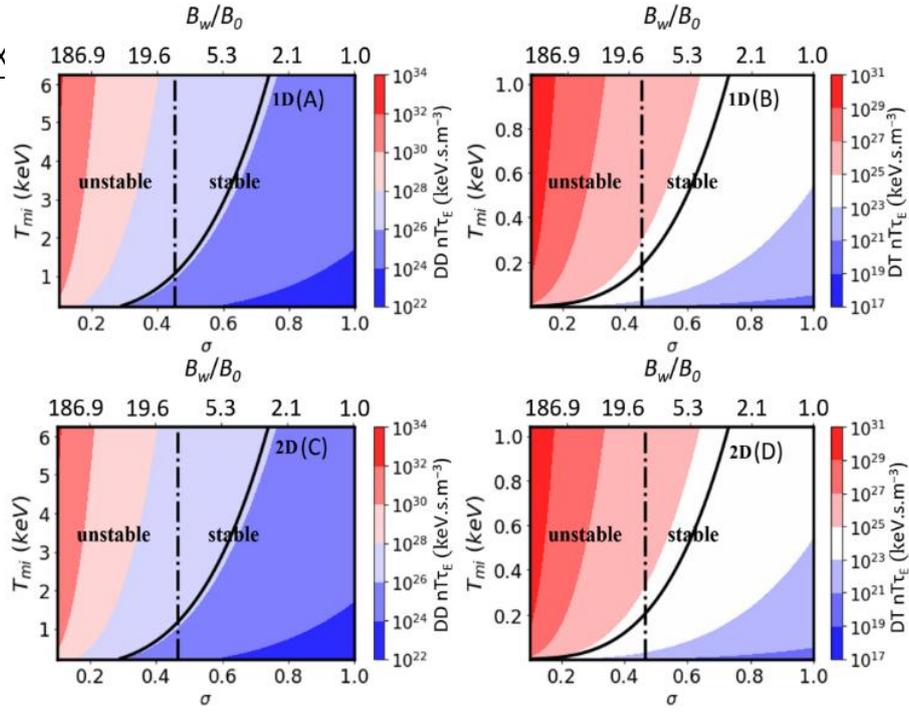

$$Y_n = \frac{1}{4}n_i^2 \langle \sigma v \rangle \pi R_s^2 l_s \tag{21}$$

**Figure 8.** Stable ignition regime from 1D scaling laws for (A) D-D reaction, (B) D-T reaction, and from 2D MHD equilibrium for (C) D-D reaction, (D) D-T reaction for different radial compression ratio $\sigma$ and magnetic field strength at the wall $B_w$, with $l_{s0} = 2.5m$, $n_{i0} = 3.0 \times 10^{19} m^{-3}$, and $R_w = 0.25m$. The black dash line denotes the s/κ stability criterion and the black solid curve the lowest ignition contour line.

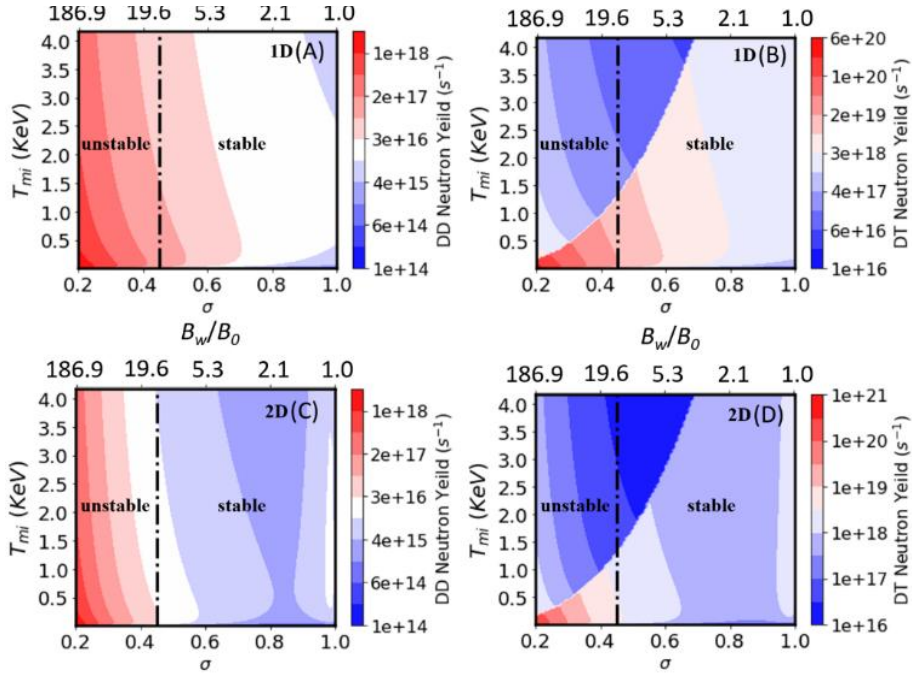

**Figure 9.** Estimates of the neutron yield rates limits from 1D scaling laws for (A) D-D reaction, (B) D-T reaction, and from 2D MHD equilibrium for (C) D-D reaction, (D) D-T reaction for different radial compression ratio $\sigma$ and magnetic field strength at the wall $B_w$, with $l_{s0} = 2.5m$, $n_{i0} = 3.0 \times 10^{19} m^{-3}$, and $R_w = 0.25m$. The black dash line denotes the s/κ stability criterion.

where $n_i$ is the ion number density, $\langle\sigma v\rangle$ is the reaction rate, $R_s = \sigma R_w$ is the plasma radius at the FRC separatrix, $l_s$ the plasma length at the separatrix. For D-D reactions, the reaction rate

$$\langle\sigma v\rangle = 2.6 \times 10^{-14} T_i^{-2/3} exp(-18.76 T_i^{-1/3}) \quad (22)$$

where $T_i$ is in keV. For D-T reactions and for $T_i < 10 keV$

$$\langle\sigma v\rangle = 3.8 \times 10^{-12} T_i^{-2/3} exp(-19.02 T_i^{-1/3}), (23)$$

for $T_i > 10 keV$

$$\langle\sigma v\rangle = 3.41 \times 10^{-14} T_i^{-2/3} exp\left(-27.217 T_i^{-\frac{2}{3}} + 3.638 T_i^{-\frac{1}{3}}\right) \quad (24)$$

Using the scalings for $n_i$ and $l_s$ amended by the MHD equilibrium, we compare the neutron yield rates for different compression ratios and magnetic fields strength at the wall in the MHD stable regimes between D-D and D-T reactions (Fig. 9). The neutron irradiation power can be calculated from $P_{N-DD} = \frac{Y_N}{\pi R_w^2 l} \times 2.45 MeV$ and $P_{N-DT} = \frac{Y_N}{\pi R_w^2 l} \times 14 MeV$ for D-D and D-T reactions respectively, where $Y_N$ is the neutron yield, $l = 2.5m$ is the length of the cylinder, $R_w = 0.25m$ is the wall radius, $2.45 MeV$ and $14 MeV$ are the neutron energy for D-D and D-T reactions respectively. Thus, the neutron irradiation power from the 1D scaling law for D-D and D-T reactions is $0.20 MW/m^2$ and $2.28 MW/m^2$ when $\sigma = 0.5, T_{mi} = 0.5 keV$ and $\sigma = 0.5, T_{mi} = 1.75 keV$ respectively. Similarly, for the 2D MHD equilibrium, the neutron irradiation power for D-D and D-T reactions is $0.03 MW/m^2$ and $0.46 MW/m^2$ when $\sigma = 0.5, T_{mi} = 0.5 keV$ and $\sigma = 0.8, T_{mi} = 2.0 keV$ respectively. The maximum neutron yield from DD-1D and DT-1D is $3 \times 10^{16} s^{-1}$ and $3 \times 10^{18} s^{-1}$, while from DD-2D and DT-2D is $4 \times 10^{15} s^{-1}$ and $1 \times 10^{18} s^{-1}$ respectively. Moreover, in pure deuterium operation, the neutron yield and the neutron power fluence is smaller than the D-T reaction. This results from the higher number density of target ions for which only half the D-D reactions yield a neutron and due to the lower energy of the D-D neutron. The neutron yield produced are in the range of low neutron rates expected at the ITER operation as also outlined by Kovalev et al 2020 [26].

## 5. Summary and discussion

In summary, the scaling laws for the adiabatic compression of FRC based on 1D analytical theory has been amended using results from 2D MHD equilibrium calculations. In particular, the FRC elongation has been self-consistently determined from the G-S equation solution for any given radial compression ratio. The amended scaling for FRC elongation during magnetic compression is applied to the estimate of the upper limits for the radial compression ratio along with the empirical stability criterion for FRC. The stability regimes for fusion ignition and neutron yield rates from the approach of FRC compression are also evaluated. Under the combined constraints from FRC two-dimensional MHD equilibrium force balance and empirical kinetic MHD stability conditions, along with the assumption that the magnetic confinement time is governed by the resistive magnetic diffusion, it is found that the FRC plasma can access the fusion ignition parameter regime through a stable quasi-static magnetic compression process, which demonstrates the physical feasibility of quasi-static magnetic compression of FRC plasma as a potential path to achieving fusion ignition conditions. These calculations may help the design of future fusion experiments and devices based on the magnetic compression of FRC plasma.

The 2D MHD equilibrium calculation for FRC during compression in this work adopts the scaling law for the maximum pressure previously derived from 1D analytical theory. We plan to develop a more self-consistent scaling law for the maximum pressure of FRC plasma during the magnetic compression entirely from the 2D MHD equilibrium and geometry of FRC in future study.

**Acknowledgements**

This work was supported by the National Magnetic Confinement Fusion Program of China Grant No. 2017YFE0301805, the National Natural Science Foundation of China Grant No. 51821005, the Fundamental Research Funds for the Central Universities at Huazhong University of Science and Technology Grant No. 2019kfyXJJS193, and the U.S. Department of Energy Grant Nos. DE-FG02-86ER53218 and DE-SC0018001. The authors are grateful for the supports from the NIMROD team. The author Abba Alhaji Bala acknowledges the support from the Chinese Government Scholarship.